\documentclass[aps,prl,twocolumn]{revtex4}
\usepackage[latin1]{inputenc}
\usepackage{amsmath}
\usepackage{amsfonts}
\usepackage{amssymb}
\usepackage{amsthm}

\ifx\pdftexversion\undefined
\fi

\theoremstyle{remark}

\theoremstyle{definition}

\begin{document}

\title{ Local Experiments See Cosmologically Varying Constants}
\author{Douglas J. Shaw}
\email{D.Shaw@damtp.cam.ac.uk}
\affiliation{DAMTP, Centre for Mathematical Sciences, University of Cambridge,
Wilberforce Road, Cambridge CB3 0WA, UK}
\author{John D. Barrow}
\email{J.D.Barrow@damtp.cam.ac.uk}
\affiliation{DAMTP, Centre for Mathematical Sciences, University of Cambridge,
Wilberforce Road, Cambridge CB3 0WA, UK}
\date{20th December 2005}

\begin{abstract}
We describe a rigorous construction, using matched asymptotic expansions,
which establishes under very general conditions that local terrestrial and
solar-system experiments will measure the effects of varying `constants' of
Nature occurring on cosmological scales to computable precision. In
particular, `constants' driven by scalar fields will still be found to
evolve in time when observed within virialised structures like clusters,
galaxies, and planetary systems. This provides a justification for combining
cosmological and terrestrial constraints on the possible time variation of
many assumed `constants' of Nature, including the fine structure constant
and Newton's gravitation constant.

\end{abstract}

\maketitle

Investigations into the possibility that the fine structure `constant', $%
\alpha $, is slowly varying \cite{webb} combine observational evidence from
quasar spectra with indirect solar system data and laboratory constraints on
allowed variations. In so doing, most commentators invariably assume that
the local and cosmological observations are directly comparable \cite%
{reviews}. However, this is a strong assumption that requires proof. Why
should local experiments `see' variations in `constants' occurring on
cosmological scales in regions which are gravitationally unbound and
participating in the universal Hubble flow? In this letter we describe the
first rigorous demonstration of when such an assumption is justified and
identify the key inequality that must be satisfied if local variations of
`constants' are to track global variations.

Theories which introduce varying constants self-consistently into Einstein's
conception of a gravitation theory do so by deriving the `constant' $\mathbb{%
C}$ from some scalar field, $\phi $, so $\mathbb{C}\rightarrow \mathbb{C}%
(\phi )$. The variations of this scalar field contribute to the spacetime
curvature like other forms of mass-energy. They must also conserve energy
and momentum and so their dynamics are constrained by a wave equation 
\begin{equation}
\square \phi =\sum_{i,j,k}\lambda _{i}f_{j}(\phi )L_{k}(\rho ,p),
\label{gen}
\end{equation}%
where $\phi $ is associated with the variation of a `constant' $\mathbb{C}$
via a relation $\mathbb{C}=f(\phi )$; the constants $\lambda _{i}$ are
dimensionless measures of the strength of the space-time variation of $%
\mathbb{C}$, $f_{j}(\phi )$ are functions determined by the definition of $%
\phi ,$ and $L_{k}(\rho ,p)$ are some linear combinations of the density, $%
\rho $, and pressure, $p$, of the matter that is coupled to the field $\phi $%
. This form includes all the standard theories for varying constants, like
those for the Newtonian gravitation 'constant' $G$, $\alpha $, and the
electron-proton mass ratio, described in refs. \cite{bd, bek, bsbm, bmmu}.

We shall refer to our scalar field as the `dilaton', denote it by $\phi $,
and analyse the form of (\ref{gen}) further by assuming that $\phi (\vec{x}%
,t)$ satisfies a conservation equation that can be reduced to the form 
\begin{equation}
\square \phi =B_{,\phi }(\phi )\kappa T-V_{,\phi }\left( \phi \right) 
\label{cons}
\end{equation}%
where $T$ is the trace of the energy momentum tensor, $T=T_{\mu }^{\mu }$,
with any contribution from a cosmological constant, $\Lambda $, excluded; we
absorb any dilaton to cosmological constant coupling into the definition of $%
V(\phi )$. The dilaton to matter coupling, $B(\phi ),$ and the
self-interaction potential, $V(\phi )$, are arbitrary functions of $\phi $, $%
\kappa \equiv 8\pi G$ and $c\equiv \hslash \equiv 1$. This covers a wide
range of theories which describe the spacetime variation of `constants' of
Nature, and includes Einstein-frame Brans-Dicke (BD), \cite{bd}, and all
other, single field, scalar-tensor theories of gravity. In cosmologies that
are composed of dust, cosmological constant and radiation it will also
contain the Bekenstein-Sandvik-Barrow-Magueijo (BSBM) of varying $\alpha $, 
\cite{bsbm}, and other (single dilaton) theories which describe the
variation of standard model couplings, \cite{posp}. We assume that the
background metric is of Friedmann-Robertson-Walker (FRW) form 
\begin{equation}
\mathrm{d}s^{2}=\mathrm{d}t^{2}-\frac{a^{2}(t)\mathrm{d}r^{2}}{%
1-kr^{2}}-a^{2}(t)r^{2}\{\mathrm{d}\theta ^{2}+\sin ^{2}\theta \mathrm{d}\phi
^{2}\} ,  \label{frw}
\end{equation}%
where $k$ is the curvature parameter. In the background universe the dilaton
is also assumed to be homogeneous (so $\phi =\phi _{c}(t),$ where $_{c}$
denotes background quantities) and satisfies: 
\begin{equation}
\frac{1}{a^{3}}\left( a^{3}\dot{\phi}_{c}\right) ^{\cdot }=B_{,\phi }(\phi
_{c})\kappa \left( \epsilon _{c}-3p_{c}\right) -V_{,\phi }\left( \phi
_{c}\right) .  \label{dilcons}
\end{equation}

\ \ The dilaton conservation equation also reduces to an ordinary
differential equation if we are only interested in its time-independent mode 
$\phi (r$) in a static spherically symmetric Schwarzschild spacetime. In the
actual universe, however, spacetimes that look static in some locality must
match smoothly on to the cosmological background on large scales. Hence, to
model the evolution of the dilaton accurately in some inhomogeneous region,
embedded in a homogeneous background, we require that at large distances $%
\phi \rightarrow \phi _{c}$ appropriately. Even if spherical symmetry is
assumed, the conservation equation is difficult to solve, either exactly or
approximately. Even numerical models are technically difficult to set-up,
see ref. \cite{harada}, and only allow us to consider one particular choice
of $B(\phi )$, $V(\phi ),$ and the spacetime background at a time. If we are
to bring all the experimental evidence to bear on these models, we need to
know when we are allowed to use local observations to constrain cosmological
variations of `constants' like $\alpha $. Hence, we would like to find the
conditions which ensure that the local time-variation of $\phi $ tracks the
cosmological one%
\begin{equation}
\dot{\phi}(\vec{x},t)\approx \dot{\phi}_{c}(t)  \label{wettcond}
\end{equation}%
to some specified precision.

We shall assume that the dilaton field is only weakly coupled to gravity,
and has a negligible effect on the background spacetime geometry. If we
ignore this back-reaction then we can consider the dilaton evolution on a
fixed background with given matter density. We will take the background to
be an exact solution to Einstein's equations with matter possessing the
following properties. The local metric is approximately Schwarzschild, with
mass $m$ at time $t=t_{0}$, inside a closed region of spacetime outside a
surface at $r=R_{s}$. The metric for $r<R_{s}$ is left unspecified.
Asymptotically, the metric must approach FRW and the whole spacetime should
become exact FRW in the limit $m\rightarrow 0$. As the local inhomogeneous
energy density of the asymptotically FRW spacetime tends to zero, the local
spacetime metric exterior to $r=R_{s}$ must tend to the Schwarzschild
metric. Similar scenarios have been considered by Wetterich in \cite%
{wetterich:2002} and in a very recent work by Avelino et al \cite{avelino}.
Whilst our conclusions will be similar to the ones claimed in both of these
studies, we believe that neither study proves what is claimed. In \cite%
{shawbarrow1} we identified an flaw in \cite{wetterich:2002}. When this is
corrected the proof provided breaks down. A key point of the matched
asymptotic expansion method is that the excluded terms in the asymptotic
expansions for $\phi $ and $\dot{\phi}$ must be shown to be smaller than the
included ones. These are non-trivial requirements. In \cite{avelino} they
are simply assumed and so no proof of the main claim is provided.

In the local, approximately Schwarzschild region, the intrinsic interior
length scale, $L_{I}$, of a sphere centred on the Schwarzschild mass with
surface area $4\pi R_{s}^{2}$ is defined by the Riemann invariant: 
\begin{equation}
L_{I}=\left( \tfrac{1}{12}R_{abcd}R^{abcd}\right) ^{-1/4}=\frac{R_{s}^{3/2}}{%
\left( 2m\right) ^{1/2}}.  \label{invar}
\end{equation}%
In the asymptotically FRW region, the intrinsic length scale is proportional
to the inverse root of the local energy density: $1/\sqrt{\kappa \epsilon
_{c}+\Lambda }$, where $\epsilon _{c}$ is the total energy density of
matter, We shall assume that the FRW region is approximately flat ($k=0$),
and we define a length scale appropriate for this exterior region at epoch
at $t=t_{0}$ equal to the inverse Hubble parameter at that time: 
\begin{equation*}
L_{E}=1/H_{0}.
\end{equation*}%
For realistic models $L_{E}\gg L_{I}$; we therefore define $\delta $ to be a
small parameter given by $\delta =L_{I}/L_{E}$.

We solve the dilaton equation (\ref{dilcons}) using an asymptotic expansion
for small $\delta $ in the interior region, where $%
L_{I}^{-1}(t-t_{0}),R_{s}^{-1}r\sim \mathcal{O}(1)$, and also in the
exterior where $H_{0}t\sim \mathcal{O}(1)\sim H_{0}r$. In general, an
asymptotic expansion valid in one region will \textit{not} also be valid in
the other. We can only enforce a subset of our boundary conditions in each
region, hence the resulting expansions will each contain undefined constants
of integration. We have used the method of matched asymptotic expansions to
remove this ambiguity in the expansions by assuming that there exists an
intermediate region, where $L_{I}^{-1}(t-t_{0}),R_{s}^{-1}r\sim \mathcal{O}%
(\delta ^{-\alpha})$ and $H_{0}t,H_{0}r\sim \mathcal{O}(\delta ^{1-\alpha})$, in which
both the interior and exterior expansions for $\phi $ remain valid. Then, by
the uniqueness of asymptotic expansions, the two expressions must be equal
there, see \cite{hinch}. In this way we can determine the previous unknown
constants of integration in each expansion, and effectively apply \emph{all}
of our boundary conditions to each expansion.

Formally, we require that, as $\delta \rightarrow 0,$ our choice of
spacetime background should be FRW at zeroth order in the exterior expansion
and Schwarzschild to lowest order in the interior expansion. All
spherically-symmetric solutions to Einstein's equations with pressureless
matter and a $\Lambda $ term fall into the Tolman-Bondi class (see ref. \cite%
{kr}). They are parametrised by two arbitrary functions of one spatial
variable, $r$. We have only looked at the flat, Tolman-Bondi models with
non-simultaneous initial singularities \cite{Gautreau:1984, kr}, and the
non-flat Tolman-Bondi models with simultaneous initial singularities. These
two subcases are fully specified by prescribing only one function: the
matter density $\epsilon (r,t_{i})$ on some initial space-like hypersurface, 
$t_{i}$. Another metric used to study the effect of the universe's expansion
on solar system dynamics is the McVittie metric \cite{mcv, kr}. In the
exterior limit the McVittie metric asymptotes to a dust-plus-$\Lambda $ FRW
cosmology. For radii $r$ where $H_{0}^{-1}\gg r\gg 2m$, the McVittie metric
looks like Schwarzschild; the horizon in the McVittie metric possesses a
naked curvature singularity and so cannot model a black hole in an expanding
universe.

To zeroth order in both the exterior and interior, all these background
choices will look alike and the solution for $\phi $ will also be the same
in all cases. As boundary conditions we take $\phi \rightarrow \phi _{c}(t)$
as $r\rightarrow \infty $, where $\phi _{c}(t)$ satisfies the cosmological
dilaton equation (\ref{dilcons}). We assume that the dilaton flux out of the
`star', with surface at $R=R_{s}$, is given, to leading order, by 
\begin{eqnarray*}
-R_{s}^{2}\left( 1-\frac{2m}{R_{s}}\right) \left. \partial _{R}\phi
_{0}\right\vert _{R=R_{s}}=2mF\left( \bar{\phi}_{0}\right)&& \\ \notag =\int_{0}^{R_{s}}%
\mathrm{d}R^{\prime }R^{\prime }{}^{2}B_{,\phi }(\phi _{0}(R^{\prime
},t))\kappa \epsilon (R^{\prime }),&&
\end{eqnarray*}%
where $R$ is the `physical radius', i.e. a surface $(t,R)=const$ has surface
area $4\pi R^{2}$. For black-holes we must demand $F\left( \bar{\phi}%
_{0}\right) =0$, we otherwise expect $F\left( \bar{\phi}_{0}\right) \approx
B_{,\phi }(\phi _{c})$. At higher orders we find the flux by perturbing the
above expression as explained in \cite{shawbarrow1}. The zeroth-order
exterior solution is then just $\phi =\phi _{c}(t)$. In the interior we find 
\begin{equation*}
\phi^{(0)}=\phi _{e}\left( \delta T\right) -F\left( \bar{\phi}_{0}\right) \ln
\left( 1-\frac{2m}{R_{s}\xi }\right) ,
\end{equation*}%
where $T=L_{I}^{-1}(t-t_{0})$ and $\xi =R_{s}^{-1}R$. The matching procedure
gives $\phi _{e}\left( \delta T\right) =\phi _{c}(t)$.

We have performed our analysis under the assumption that the background
universe is described by McVittie metric. In isotropic coordinates this is 
\begin{eqnarray}
\mathrm{d}s^{2}=\left[ \frac{1-\mu (t,r)}{1+\mu (t,r)}\right] ^{2}dt^{2}-%
\frac{[1+\mu (t,r)]^{4}}{(1+\tfrac{1}{4}kr^{2})^{2}}a^{2}(t)\left[ \mathrm{d}%
r^{2}\right.&& \\ \notag \left. +r^{2}\{\mathrm{d}\theta ^{2}+\sin ^{2}\theta \mathrm{d}\phi ^{2}\}%
\right],&&  \label{mcv1}
\end{eqnarray}%
where 
\begin{equation}
\mu (t,r)=\frac{m}{2ra(t)}\left( 1+\tfrac{1}{4}kr^{2}\right) ^{1/2},
\label{mcv2}
\end{equation}%
and $m$ and $k$ are arbitrary constants. In the limit $a=1$, $m$ is the
Schwarzschild mass, $k$ is the curvature of the surfaces $(t,r)=\mathrm{const%
}$ in the $m=0$ (i.e. FRW) limit. The McVittie metric provides a good
approximation to the exterior metric of a massive spherical body with a
physical radius much larger than its Schwarzschild radius. In the interior
it is most convenient to work with%
\begin{equation*}
R=\left[ 1+\mu \right] (1+\tfrac{1}{4}kr^{2})^{-1/2}a(t)r
\end{equation*}%
as the radial coordinate. If $k=0$, then the surface $(t,R)=\mathrm{const}$
has area $4\pi R^{2}$. When $t=t_{0}$ the interior geometry will be
approximately Schwarzschild over scales where $\mathrm{d}R\sim \mathcal{O}%
(R_{s})$ and $\mathrm{d}t\sim \mathcal{O}(L_{I})$. We define new
dimensionless coordinates by 
\begin{equation*}
T=(t-t_{0})/L_{I}\text{ and }\xi =R/R_{s}.
\end{equation*}%
In ref. \cite{shawbarrow1} we found the interior solution for the dilaton
field to $\mathcal{O}(\delta ^{2})$ and the exterior solution to order $%
\mathcal{O}(\delta )$, and explicitly performed the matching. The inner
approximation to $\phi $ is then given by: 
\begin{equation*}
\phi _{I}\sim \phi ^{(0)}+\delta ^{2}\frac{2m}{R_{s}}\phi _{I}^{(1)}+%
\mathcal{O}\left( \delta ^{4}\right) ,
\end{equation*}%
\noindent where 
\begin{widetext}
\begin{eqnarray}
\phi _{I}^{(1)} &\sim &\frac{2m\xi }{R_{s}}\left( \tfrac{1}{2}\left( \phi
_{E}^{(0)\prime \prime }+h\phi _{E}^{(0)\prime }-h^{\prime }F\left( \bar{\phi%
}_{0}\right) \right) +\tfrac{1}{8}\Omega _{k}(\delta T)F\left( \bar{\phi}%
_{0}\right) -h^{2}F\left( \bar{\phi}_{0}\right) +\frac{3}{2}B_{,\phi }(\phi _{E})\left( h^{\prime }(\delta T)+%
\tfrac{11}{4}\Omega _{k}(\delta T)\right)  \right.   \label{mcintexp} \\
&& \left. -\frac{1}{2}F\left( \bar{\phi}_{0}\right) B_{,\phi \phi }\left( \phi _{E}\right) R_{s}^{2}\kappa \epsilon
_{dust}^{(0)}+F\left( \bar{\phi}_{0}\right) R_{s}^{2}V_{,\phi \phi }\left(
\phi _{E}\right) \right) +\frac{2m}{R_{s}}\left( \frac{C(\delta T)}{\xi }+D(\delta T)\right) +%
\mathcal{O}\left( \frac{2m}{R_{s}}\right) ^{2},  \notag
\end{eqnarray}
\end{widetext}
\noindent $\phi _{c}(t)=\phi _{E}^{(0)}(\tau )$ is the cosmological value of
dilaton field, and $C(\delta T)$ and $F\left( \bar{\phi}_{0}\right) $ are
`constants' of integration to be specified by a boundary condition at $%
R=R_{s}$ depending on the details of the energy-distribution in $R<R_{s}$.
We do not expect these two latter terms to be any larger than the other
terms in the above expression when $\xi =1$. Our primary concern is the
behaviour of the time derivative of $\phi $ in the interior and in the
McVittie background. We have 
\begin{equation*}
\frac{\partial _{t}\phi _{I}}{\partial _{t}\phi _{c}}\sim 1-F_{,\phi }\left( 
\bar{\phi}_{0}\right) \ln \left( 1-\frac{2m}{R}\right) +\mathcal{O}\left(
\delta ^{2}\left( \frac{2m}{R}\right) ^{2}\right) .
\end{equation*}%
Since, in most cases of interest, the dilaton to matter coupling is weak and
we are far outside the Schwarzschild radius of the Sun, both $F(\left( \bar{%
\phi}_{0}\right) $ and $2m/R$ are $\ll 1$. Hence, $\partial _{t}\phi
_{I}/\partial _{t}\phi _{c}\approx 1$, and the local time evolution of the
dilaton tracks the cosmological one. The strength of this result arises
partly from our restrictive choice of background metric.

We now consider Tolman-Bondi backgrounds, where more interesting behaviour
is allowed. The background metric is 
\begin{equation*}
\mathrm{d}s^{2}=\mathrm{d}t^{2}-\frac{R_{,r}^{2}(t,r)}{1-k(r)}\mathrm{d}%
r^{2}-R^{2}(t,r)\{\mathrm{d}\theta ^{2}+\sin ^{2}\theta \mathrm{d}\phi
^{2}\},
\end{equation*}%
where 
\begin{equation*}
R_{,t}^{2}=-k(r)+\frac{2m+2Z\left( r\right) }{R}+\frac{1}{3}\Lambda R^{2}.
\end{equation*}%
The matter content of these models is pressureless dust and there is a
cosmological constant, $\Lambda $. To reproduce our required geometrical
set-up, with the dust density positive everywhere, and approach to spatial
homogeneity at large $r$, we need 
\begin{eqnarray}
Z(R_{s}) =0,\quad Z_{,r}\geqslant 0,  \quad \lim_{r\rightarrow \infty }Z(r) &=&\frac{1}{2}\Omega
_{dust}^{(0)}H_{0}^{2}r^{3}.  \notag
\end{eqnarray}%
In the interior we take $T=(t-t_{0})/L_{I}$, and $\xi =R/R_{s}$ as
coordinates and 
\begin{equation*}
Z(r)/m\sim \delta ^{q}\mu _{1}(\eta )+o\left( \delta ^{q}\right) ,
\end{equation*}%
where $\eta =\left( \xi ^{3/2}+3T/2\right) ^{2/3}$ and $R_{s}\eta =r+%
\mathcal{O}(\delta ^{q},\delta ^{2/3})$. From the exact solutions (with $r=R$
at $t=t_{0}$) we find 
\begin{equation*}
k(\eta )=\delta ^{2/3}k_{0}\left( 1+\delta ^{q}\mu _{1}(\eta )+o\left(
\delta ^{q}\right) \right) +\mathcal{O}\left( \delta ^{5/3}\right) ,
\end{equation*}%
where $k_{0}=({2m}/{R_{s}})( \pi / H_{0}t) ^{2/3}.$
Solving for $\phi $ in the interior to order $\delta ^{q}$ we find $$\phi _{I} \sim \phi^{(0)}_I(\delta T)+\delta ^{q}\phi _{I}^{(1)}+o(\delta ^{q}),$$ where $\phi^{(0)}_I$ is given by
\begin{eqnarray}
\phi ^{(0)}_I &=&\phi _{e}\left( \delta \sqrt{1-\delta ^{2/3}k_{0}}\tilde{T}%
^{\ast }\right) -F\left( \bar{\phi}_{0}\right) \ln \left( 1-\frac{2m}{%
R_{s}\xi }\right) , \notag
\end{eqnarray}
and the transformation $T \rightarrow \tilde{T}^{\ast}$ is given in ref. \cite{shawbarrow1} along with the expression for $\phi _{I}^{(1)}$. We find that the time variation of $\phi_{I}$ is given by
\begin{widetext}
\begin{eqnarray} 
 \label{gatphitev}
&\phi _{I,\tilde{T}\ast } &\approx \delta \sqrt{1-\delta ^{2/3}k_{0}}\phi
_{e}^{\prime }(\sqrt{1-\delta ^{2/3}k_{0}}\delta \tilde{T}^{\ast })\left(
1-F_{,\phi }\left( \bar{\phi}_{0}\right) \ln \left( 1-\frac{2m}{R_{s}\xi }%
\right) \right)   \label{ltbphitev} \\
&&-\delta ^{q}B_{,\phi }(\phi _{c})\frac{2m}{R_{s}}\left[ \int^{\eta }%
\mathrm{d}\eta ^{\prime }\frac{\mu _{1}\left( \eta ^{\prime }\right) _{,\eta
}}{\xi ^{5/2}(\eta ^{\prime },T)}+\left( 1-\frac{F(\bar{\phi}_{0})}{B(\phi
_{c})}\right) \frac{\mu _{1,\eta }(\xi =1,T^{\ast })}{\xi ^{3/2}}\right] +...
\notag
\end{eqnarray}
\end{widetext}
Here, the excluded terms are guaranteed to be smaller than the one of the
two included terms; however, since they might be larger than the other term,
both numerically and in the limit $\delta \rightarrow 0$ , the above
expression is not a formal asymptotic approximation. When the first term in
eq.(\ref{gatphitev}) dominates, condition (\ref{wettcond}) holds, and the
local time evolution of the scalar field is, to leading order, the same as
the cosmological one. Assuming our background choice is suitable for the
application of this method, condition (\ref{wettcond}) fails to hold if, and
only if, the second term in this expression dominates over the first.

We have explicitly checked that the McVittie background matching works, and
by analysing the forms of the interior and exterior expansions in the
Tolman-Bondi metric we can show that the matching procedure will break down
if $m<1-p\alpha \leq 1-p(1-m)^{-1}$or if $m\geq 1-p\alpha $ when $p\alpha =0$%
, where $\mu _{1}\left( \chi \right) \sim \chi ^{n}$ as $\chi \rightarrow
\infty $ for some $n>0$, and where $Z=Z_{FRW}+\delta ^{p}z_{1}+\mathcal{O}%
(\delta ^{q})$, and $z_{1}(H_{0}r)\sim (H_{0}r)^{m}$. By considering the
behaviour of the next-to-leading order terms we can say that if both the
interior and exterior zeroth-order approximations are simultaneously valid
in some intermediate scaling region we need 
\begin{equation*}
\lim_{\delta \rightarrow 0}\left( R^{2}\kappa \epsilon (R,t)\right) =o(1)
\end{equation*}%
for all $\alpha \in (0,1)$, where the limit is taken with $\{L_{I}^{\alpha
}L_{E}^{1-\alpha }(t-t_{0}),L_{I}^{\alpha }L_{E}^{1-\alpha }R\}$ held fixed.
This is also the condition for matching that we have found in the
non-spherically symmetric extension of our result using the Szekeres
spacetime \cite{szekeres}. If such an intermediate region does indeed exist
then the zeroth-order matching will be valid. The general form of the
interior approximation to $\mathcal{O}\left( \delta ^{p}\right) $ will then
be correct; the only unknown function in it is $B(\delta T)$. If the
matching works to order $\delta ^{p}$, as well as zeroth-order, then we have
argued in \cite{shawbarrow1} that $B(\delta T)$ will be quasi-static. If the
matching procedure does not work to this order then its quasi-static
character may be lost. However, we would not expect it to vary in time any
faster than the other $\mathcal{O}\left( \delta ^{p}\right) $ terms. So long
as we can match the zeroth-order approximations in some region, we can find
the circumstances under which condition (\ref{wettcond}) holds by comparing
the sizes of the two terms in (\ref{gatphitev}). By adding sub-leading order
terms to this condition so as to express it in terms of physical quantities,
we find \cite{shawbarrow1} that to leading order it is equivalent to 
\begin{eqnarray}
I:=&&\frac{\dot{\phi}-\dot{\phi}_{c}}{\phi _{c}} \label{simpform} =\frac{B_{,\phi }(\phi
_{c})\int_{\gamma (R)}\mathrm{d}l\,\Delta (R_{,t}\kappa \epsilon )}{\dot{\phi%
}_{c}(t)}\\ \notag &&-\int_{\gamma (R)}\mathrm{d}l\Delta (R_{,t}) \frac{\ddot{\phi}_c(t)}{\dot{\phi}_c(t)} \ll 1  
\end{eqnarray}%
where $\mathrm{d}l=\mathrm{d}r^{\prime }R_{,r}$, and the path of
integration, $\gamma (R)$, runs from $R$ to $\infty $, and $\Delta
(R_{,t}\epsilon )=R_{,t}\kappa {\epsilon }-HR\kappa \epsilon _{c}(t)$, $%
\Delta (R_{,t})=R_{,t}-HR$. $R_{,t}(r,t)$ is instantaneous the velocity of a
matter particle at $(r,t)$. Since we expect $\phi $ will only feel its
causal past, we should take $\gamma (R)$ to run from $R$ to spatial infinity
along a past radially-directed light-ray. Eq. (\ref{simpform}) has the form
expected for condition (\ref{wettcond}) to hold inside a local,
spherically-symmetric, inhomogeneous region produced by a Schwarzschild mass
embedded in an asymptotically FRW universe: the LHS of eq. (\ref{simpform})
vanishes as we approach the cosmological region and the local time variation
of the dilaton field is driven by the time variation of the energy density
along its past light-cone. The second, or \emph{drag,} term in \ref{simpform}
comes from the change of time-coordinate $T\rightarrow \tilde{T}^{\ast }$
and is small compared to the first term whenever there is a local
overdensity of matter.

Therefore we are led to conjecture that eq.(\ref{simpform}) is a general
sufficient condition for eq.(\ref{wettcond}) to hold: applicable even in
spherically-symmetric (or almost spherically-symmetric), dust plus
cosmological constant, backgrounds in which the matching procedure formally
fails.

If the cosmic evolution of the dilaton field is dominated by its coupling to
matter, so $\left\vert B_{,\phi }(\phi _{c})\kappa \epsilon _{c}\right\vert
\gg \left\vert V_{,phi}(\phi _{c})\right\vert $, then $\dot{\phi}_{c}(t)=%
\mathcal{O}\left( B_{,\phi }(\phi _{c})\kappa \epsilon _{c}(t)/H(t)\right) $
and so condition (\ref{wettcond}) holds at the time $t=t_{0}$ whenever 
\begin{equation}
\int_{\gamma (R)}\mathrm{d}l\,H(t_{0})\,\frac{\left\vert \Delta
(R_{,t}\epsilon )\right\vert }{\epsilon _{c}(t_{0})}\ll 1.  \label{simpform2}
\end{equation}%
where we assumed the \emph{drag} term to be negligible and dropped it.
Alternatively, if the cosmic evolution is potential-driven, so $\left\vert
V_{,\phi }(\phi _{c})\right\vert \gg \left\vert B_{,\phi }(\phi _{c})\kappa
\epsilon _{c}\right\vert $, then the LHS of the above expression will be
suppressed by an additional factor of $\left\vert B_{,\phi }(\phi
_{c})\kappa \epsilon _{c}/V_{,\phi }(\phi _{c})\right\vert \ll 1$.

We can apply out results to solve our original problem: to determine if
condition (\ref{wettcond}) holds when the local region is the Earth or the
solar system. Consider a star (and associated planetary system) inside a
galaxy that is embedded in a large cluster. The cluster, of size $R_{clust}$%
, is assumed to have virialised. We consider the different contributions to
the LHS of condition (\ref{simpform2}) in this astronomical hierarchy, with: $I=I_{clust}+I_{gal}+I_{star}$.

Since we have only performed our calculations for spherically-symmetric
backgrounds, we should, strictly speaking, also require spherical symmetry
about our star. However, in a forthcoming paper we will show, by determining
the explicit matched asymptotic expansions for the case of the Szekeres
background metric \cite{szekeres}, which possesses no Killing vectors \cite%
{bonn}, that criterion (\ref{simpform}) still gives a very good
approximation to the true condition in the absence of spherical symmetry.
For illustration we have evaluated $I$ for Brans-Dicke theory. For other
theories, $I$ will still take very similar values. We assume that the
density of non-virialised matter, just outside the virialised cluster, is no
greater than the average density of the cluster and if we move away from the
edge of the virialised region, the over-density drops off quickly, i.e. as $%
R^{-s}$, $s>1/2$. In this case, the Tolman-Bondi result, eq. (\ref{ltbphitev}%
), applies and the magnitude of $I$ is bounded by: 
\begin{eqnarray*}
I_{clust} &\lesssim &t_{0}^{-1}(s-1/2)^{-1}\sqrt{2M_{clust}R_{clust}}\frac{%
\epsilon _{clust}}{\epsilon _{c}} \\ \notag &\approx& \frac{1.2\times 10^{-7}}{\Omega_{m}(s-1/2)}\left(\frac{R_{clust}}{\mathrm{Mpc}}\right)\left(\frac{v_{clust}}{\mathrm{km}\,\mathrm{s}^{-1}}\right)(1+z_{vir})^{3} \\
&\approx &1.5 \times 10^{-3}[3/(2s-1)]\Omega _{m}^{-1}(1+z_{vir})^{3}\ll 1,
\end{eqnarray*}%
\noindent where we have used $3GM/R_{clust}=v_{clust}^{2}$; $R_{clust}$ is
the post-virialisation scale of the cluster, and $v_{clust}$ is its average
virial velocity; $t_{0}=13.7Gyrs$ is the age of the universe. In the final
approximation we used representative values $R_{clust}=100Mpc$ and $%
v_{clust}=200\mathrm{km\,s}^{-1}$ appropriate for a cluster like Coma.
Taking a cosmological density parameter equal to $\Omega _{m}=0.27,$ in
accordance with WMAP we expect that for a typical cluster which virialised
at a redshift $z_{vir}\ll 1$, we would have $I_{clust}\approx 5.7\times
10^{-3}$. The term in $[]$ is unity when $s=2$, i.e. $2GM/r \rightarrow const$; such a matter distribution is characteristic of dark matter halos. Different
choices of $s>1/2$ can be seen to only change this estimate by an $\mathcal{O%
}(1)$ factor. As $s\rightarrow 1/2$ the matched asymptotic expansion method,
and hence this particular evaluation, breaks down. We believe that the
generalised formula for $I$ will, however, still give accurate results. For
the star and galaxy contribution, we assume each is accreting matter from
the interstellar medium (ISM) and intergalactic medium (IGM) respectively.
We find 
\begin{eqnarray*}
I_{gal} &\approx &-2t_{0}^{-1}\sqrt{2GM_{gal}R_{G}^{(gal)}}\frac{%
\epsilon _{IGM}}{\epsilon _{c}} \\
&= &-2.4\times 10^{-15}\Omega _{m}^{-1}\frac{M_{gal}}{M_{\odot }}%
(1+z_{vir})^{3}\left( \frac{\mathrm{km\,s}^{-1}}{v_{clust}}\right) \\ \notag &\approx
&-4.4 \times 10^{-6}(1+z_{vir})^{3}\ll 1, 
\end{eqnarray*}
\begin{eqnarray*}
I_{star} &\approx &-2t_{0}^{-1}\sqrt{2Gmr_{G}}\frac{\epsilon _{ISM}%
}{\epsilon _{c}} \\ \notag &=&-1.2\times 10^{-13}\Omega _{m}^{-1}\frac{m}{M_{\odot }}%
\left( \frac{\mathrm{km\,s}^{-1}}{v_{ISM}}\right) h^{-2}n \\
&\approx &-1.8\times 10^{-14}n\ll 1,
\end{eqnarray*}%
\noindent where we have taken $\epsilon _{IGM}\approx \epsilon _{clust}$ and 
$v_{IGM}\approx v_{clust}$, $M_{gal}=10^{12}M_{\odot }$, $v_{clust}=200%
\mathrm{km\,s}^{-1}$, $R_{G}=2GM_{gal}/v_{IGM}^{2}$, $r_{G}=2Gm/v_{ISM}^{2}$%
, and $\epsilon _{ISM}=n$ $\mathrm{protons\,cm^{-3},}$ where we expect $%
n\approx 1-10^{4}$, using $v_{ISM}=5\mathrm{km\,s}^{-1}$ and $m=M_{\odot }$
for the numerical estimates. We take $h\equiv
(H_{0}/100kms^{-1}Mpc^{-1})=0.71$ for the Hubble constant. It is clear that,
in general, $|I_{star}|\ll |I_{gal}|\ll |I_{clust}|$ as expected. The infall
of dust into the cluster will tend to be the dominant contribution to the
LHS of eq. (\ref{simpform2}). Our estimate of $I_{clust}$ should be viewed
as an upper bound on its value, which is very small compared with unity and
hence we expect condition (\ref{wettcond}) to hold near the Sun; with
deviations from that behaviour bounded by the 0.6\% level (if the dilaton
couples only to baryonic matter this could rise to as high as 3\%). Assuming
that the conditions in our solar system are not too different from those
considered above, we conclude that irrespective of the value of the
dilaton-to-matter coupling, and what dominates the cosmic dilaton evolution,
that the condition (\ref{wettcond}) will hold in the solar system in
general, and on Earth in particular, to high precision. Cosmological
variations of 'constants', like $\alpha $, will therefore be detected
directly in terrestrial experiments to the accuracy specified by the value
of $I$.

We can also conclude that, even if condition (\ref{wettcond}) is violated
during the collapse of an overdense region of matter, then once the region
stops collapsing as a result of its virialisation, the time evolution of the
dilaton field continue evolving towards homogeneity. Contrary to what has
been assumed before in the literature, local virialisation does \emph{not}
stabilise the value of the dilaton, and protect it from any global
cosmological variation. In addition, by considering what happens when $%
2m/R_{s}\rightarrow 0$, we show in ref. \cite{shawbarrow1} that we expect
the time variation of $\phi $ (and hence $G$ in a scalar-tensor gravity
theory like Brans-Dicke \cite{mem1}) on the horizon of a black-hole (as
measured by an infalling observer) to track its cosmological value,
confirming the prediction made by Jacobson, \cite{jacobson}, the
inhomogeneous models of ref. \cite{mem2}, and the numerical calculations of
Harada et. al., \cite{harada}, who also studied the Tolman-Bondi background.
We have found that whenever weak condition, $\epsilon _{l}/\epsilon _{c}\ll
1/(2mH_{0})(\gg 1)$ holds, that Jacobson's particular solution to $\square
\phi =0$ in the Schwarzschild metric is a true asymptotic approximation to
the local behaviour of the dilaton close to the Schwarzschild horizon.

In conclusion: we have used the method of matched asymptotic expansions to
find a sufficient condition for the local time-variation of a scalar field,
or varying physical `constant' driven by such a field, to track the
cosmological evolution, and we have proposed a generalisation of this
condition that is applicable to scenarios more general than those explicitly
considered here. By numerically evaluating the condition for an astronomical
scenario similar to the one appropriate for our solar system, we have
concluded that almost irrespective of the form of the dilaton-to-matter, and
the form of the dilaton self-interaction, its time variation in the solar
system and in terrestrial laboratories will track the cosmological one. We
have therefore provided a general proof of what was previously merely
assumed: that \emph{terrestrial} and \emph{solar system} based observations
can legitimately be used to constrain the \emph{cosmological} time variation
of many supposed `constants' of Nature.

\begin{acknowledgments}
We would like to thank Timothy Clifton and Peter D'Eath for discussions. DS
is supported by a PPARC studentship.
\end{acknowledgments}

\end{document}